\documentclass[sigchi-a]{acmart}

\def\BibTeX{{\rm B\kern-.05em{\sc i\kern-.025em b}\kern-.08emT\kern-.1667em\lower.7ex\hbox{E}\kern-.125emX}}
    
\copyrightyear{2019}
\acmYear{2019}
\setcopyright{none}
\acmConference[Computational Modeling in HCI: ACM CHI 2019 Workshop]{Computational Modeling in Human--Computer Interaction: ACM CHI 2019 Workshop}{May 5, 2019}{Glasgow, UK}
\acmBooktitle{Computational Modeling in Human--Computer Interaction: ACM CHI 2019 Workshop, May 5, 2019, Glasgow, UK}
\acmPrice{}
\acmDOI{}
\acmISBN{}
\begin{document}

\title{Interactive AI with a Theory of Mind}

\author{Mustafa Mert \c{C}elikok}
\authornote{Both authors contributed equally to this research.}
\email{firstname.lastname@aalto.fi}
\author{Tomi Peltola}
\authornotemark[1]
\author{Pedram Daee}
\author{Samuel Kaski}
\affiliation{%
  \institution{Helsinki Institute for Information Technology, Dept.\ of Computer Science, Aalto University}
  \city{Espoo}
  \country{Finland}
}

\renewcommand{\shortauthors}{\c{C}elikok and Peltola, et al.}

\begin{abstract}
Understanding each other is the key to success in collaboration. For humans, attributing mental states to others, the theory of mind, provides the crucial advantage. We argue for formulating human--AI interaction as a multi-agent problem, endowing AI with a computational theory of mind to understand and anticipate the user. To differentiate the approach from previous work, we introduce a categorisation of user modelling approaches based on the level of agency learnt in the interaction. We describe our recent work in using nested multi-agent modelling to formulate user models for multi-armed bandit based interactive AI systems, including a proof-of-concept user study.
\end{abstract}


\maketitle

\section{Introduction}

Artificial agents, constructed using deep learning and probabilistic modelling, have achieved human-level or super-human performance in specific well-defined domains, such as competitive games and character recognition and generation \cite{lake2017building}. Yet, they lack in abilities that are important in human social contexts, often involving ambiguous, context-dependent information and communication. This cripples human--AI collaboration.

Humans interact with intelligent systems every day for tasks that are growing in complexity. However, currently the AI does not understand humans and humans do not understand the AI. We argue that these two problems have the same solution and that the solution is the key to efficient human--AI collaboration: both need to form a model of the other. In human--human interaction, this is facilitated by a cognitive ability called the theory of mind. By understanding and being able to anticipate the other, one can make oneself understandable, take actions that make sense to the other, and complement their skills in collaborative tasks.

Recent developments in cognitive science indicate that human decision-making can be posed as boundedly rational reinforcement learning (RL). To endow interactive AI with a theory of mind of the user, we propose to look into computational models of theory of mind, and, in particular, multi-agent modelling. This enables progressing from passive views of users to acknowledging users as active agents exhibiting strategic behaviour. In the following, we will motivate the solution from human--human collaboration, differentiate it from previous work in user modelling, discuss the state of the art in computational multi-agent modelling, and describe a proof-of-concept user study. Theory of mind has recently also been recognised as the next grand challenge in cooperative multi-agent reinforcement learning \cite{bard2019hanabi}. We argue for its place in user modelling in human--computer interaction. 

\section{From theory of mind to AI's with theory of mind}

How is it that people from diverse backgrounds with different beliefs, mental capabilities, skills, and knowledge can come together and collaborate so effectively? Why are humans so good at working together? For cognitive scientists, these are important questions, and the answers involve the theory of mind (ToM). ToM is the human ability to attribute beliefs, desires, capabilities, goals, and mental states to others \cite{premack_woodruff_1978}. This allows humans to reason about others' thoughts and behaviours, and anticipate them. Studies have shown that the ToM capabilities of a team's members are a strong predictor of the team performance \cite{collective_intelligence_first_paper_tom, collective_intelligence_second_paper_tom}. Let it be collaborative or competitive, being able to understand others gives humans a big advantage in interacting with each other.

For intelligent systems, learning from human behaviour poses challenges. Inverse reinforcement learning (IRL) aims to infer the reward function of an expert from the data of its behaviour \cite{russell1998learning}. The main assumption is that, once the reward function (or the preferences) of the demonstrator is known, its behaviour can be imitated.  However, the learning problem behind IRL is ill-posed unless the planning algorithm of the demonstrator is taken into account, and hence reward function alone is not enough to reason about human behaviour in complex environments. Furthermore, single-agent methods are not a good fit to model the interactive nature of collaboration.

In multi-agent problems, agents which try to reason about each other have been explored by game theorists since 1960s \cite{harsanyi1967games}. Recently, multi-agent RL has started focusing on modelling others as well, with different approaches such as taking opponent's learning into account\cite{foerster2018learning}, learning representations of different agent behaviours\cite{mtom}, and modelling the decision processes of others in interaction\cite{gmytrasiewicz2005framework}.
ToM has a high impact in human-human and agent-agent interaction. It helps us understand why humans are so good at collaboration, and also develop multi-agent systems with better properties. However, we believe the results of ToM from human-human and agent-agent interaction also carry to human-AI interaction. They offer us ways to design intelligent systems that can collaborate with humans better. 

\section{Levels of agency in AI's user models}

User modelling is used to personalise user experiences and improve usability and performance of collaborative human--computer systems \cite{fischer2001user}. To facilitate efficient interaction, the computer makes assumptions and possibly learns about the goals, intentions, beliefs, behaviour, or other attributes of the user. Recently, user models have incorporated statistical and machine learning methods to automatically interpret and generalise (across users and tasks) the data collected in the interaction.

\begin{marginfigure}
\includegraphics[scale=0.52]{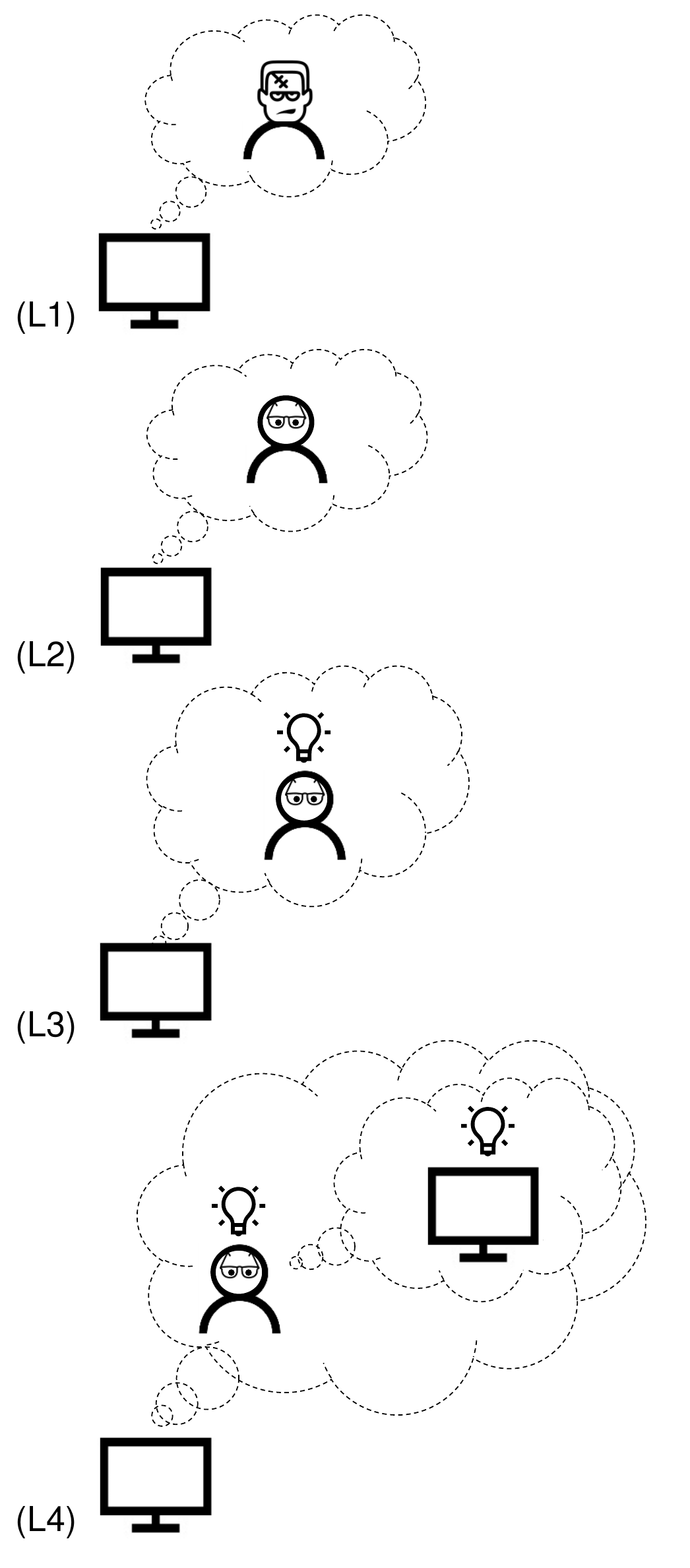}
\Description{Four level of agency in user models}
\caption{
AI's assumption of the user. Schematic of the four-level categorisation of user modelling approaches. (L1) Prescriptive and fixed model of the user. (L2) Passive user profile that is learned from interaction data. (L3) User as an active agent. (L4) User as an active agent who has a model of the system or artificial agent that adapts based on the user's actions.
}
\label{fig:user_model_types}
\end{marginfigure}

To differentiate theory-of-mind-based user modelling from previous work, we introduce a categorisation of user modelling approaches to four levels based their assumptions about the behaviour of the user during
interaction. In particular, we consider how adaptive the user is assumed to be with regard to the system, with higher levels implying more complex learning problem of the user model during the interaction (Figure~\ref{fig:user_model_types}):
\begin{enumerate}
 \item [(L1)] fixed, prescriptive behaviour,
 \item [(L2)] passive, reactive user behaviour,
 \item [(L3)] actively planning user using a passive, stationary system,
 \item [(L4)] actively planning user using an adaptive, non-stationary system.
\end{enumerate}

Level 1 prescribes a model for the user. This can include advanced and complex user models, for example, based on cognitive architectures or principles of computational rationality. In particular, fully specified, pre-trained reinforcement learning models (based on fixed, known reward function) fall into this category. Yet, L1 user models do not adapt or learn based on the interaction data and have limited means of accounting for individual variability. Level 2 models, on the other hand, learn a profile (e.g., interests, preferences, goals, type) of the user from or during the interaction, but do not otherwise model strategic behaviour. These models include collaborative filtering and multi-armed bandits, which have been highly successful in recommendation systems.

Level 3 models acknowledge the user as an active agent who plans or at least acts based on a static or stationary model of the environment. This level includes inverse reinforcement learning and theory of mind models, such as the Bayesian theory of mind \cite{baker2017rational} and machine theory of mind \cite{mtom}, that learn about and predict the agent's behaviour by observing its actions. For example, Daee et al.\ \cite{Daee2018overfitting} modelled the user as an active agent who updates her knowledge while interacting with a knowledge elicitation system. By inverting the knowledge update, the user model enables the system to extract the user's tacit knowledge. Level 4 further acknowledges that the user will form a model of the interactive system which they are using. In particular, L4 differs from L3 when the system is also adaptive and learns about the environment and the user based on observation during the interaction. The system's user model then contains a model of the system, including the adaptive behaviour, based on which the user plans. This leads to nested or recursive computational multi-agent modelling.
\begin{marginfigure}
\includegraphics[scale=1]{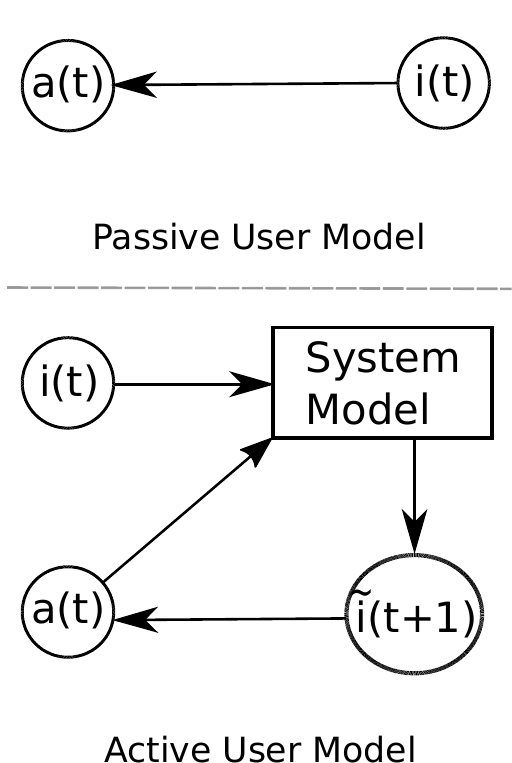}
\Description{Four level of agency in user models}
\caption{
In the passive user model (top), the feedback at time $t$, $a(t)$, depends only on the item at the time, $i(t)$. In the active user model (bottom), the user anticipates the system's next item choice $\tilde{i}(t+1)$ based on her model of the system, and chooses her action $a(t)$ towards favourable outcomes. From the system's perspective, $\tilde{i}(t+1)$ depends on $a(t)$ and $i(t)$ .
}
\label{fig:user_models_technical}
\end{marginfigure}

Which level is appropriate for an interactive system depends on the application. While most existing work in user modelling fall into levels 1 and 2, we believe that levels 3 and 4 will be important in the next generation of human--AI collaboration systems. We discuss below computational modelling approaches for implementing theory-of-mind-based user models.

\section{Computational theory of mind models}

In multi-agent literature, ToM-based models aim to capture the strategic behaviour and the underlying beliefs of others for choosing optimal actions \cite{hern2017survey}. They fall into the recursive reasoning category of opponent modelling \cite{pynadath2002communicative,gmytrasiewicz2005framework,OliehoekAmato16book,albrecht2018autonomous}. Agents are modelled in increasing levels of complexity, with higher-level agents having greater capacity to accommodate the actions of other agents in their planning. One of the earliest examples of such theory-of-mind methods is the cognitive hierarchy theory \cite{doi:10.1162/0033553041502225}. More recently, the interactive partially observable Markov decision processes (I-POMDPs) provide a decision-theoretic framework for recursive agents that act independently, and only assumes partial observability about the environment and the other agents  \cite{gmytrasiewicz2005framework}. An I-POMDP agent uses its beliefs about its opponent's model to anticipate the opponent's actions. The computational complexity of such agents is a large challenge. Recent probabilistic ToM models can be seen as specializations or extensions of this framework \cite{baker2017rational,peltola2018modelling}. A complementary line of ToM modelling in deep learning uses meta-learning to be able to anticipate actions of individual agents from sequences of observations \cite{mtom}.

In a recent work, we developed a nested multi-agent user model for multi-armed bandit based interactive systems \cite{peltola2018modelling}. In a simple exploratory search setting, the interaction rounds consist of the system suggesting an item $i(t)$ to the user and the user providing feedback $a(t)$. The system tries to help the user by aiming to maximize the cumulative reward of the item choices, $\sum_{t=1}^{T} \pi_{i(t)}$, where $\pi_{i(t)}$ is the reward for item $i(t)$. A \textbf{passive user model} would try to learn the user's preference profile, corresponding to $\pi_i$ for all $i$, from the interaction. Yet, if a user is able to anticipate the system's suggestions, they can \textbf{actively plan} their feedbacks to steer the system towards relevant items. The \textbf{active user model} is based on modelling the user as planning in a Markov decision process that nests a model of the system's bandit. The \textit{look-ahead behaviour} is inspired by the problem-space theory for human problem solving \cite{Newell:1972:HPS:1095704}. Figure~\ref{fig:user_models_technical} illustrates the differences between the user models.

\section{Proof-of-concept user study}

Consider a simple version of the Twenty Questions game where the human player selects a target word and the AI player needs to guess the word by asking sequential questions about relevance of different words to the unknown target. The human player is instructed to help the AI to find the target word, as fast as possible, by sequentially providing yes/no answers to AI's questions. The problem, from the AI point of view, is a multi-armed bandit where the word with the maximum relevance (among a fixed set of words) needs to be found with minimum number of interactions. We implemented two different user models for this task, a \textbf{passive user model} that assumes the user's answers originate from a stationary relevance profile, and an \textbf{active user model} that assumes that the user has a model of the AI she is interacting with.   

\begin{marginfigure}
\includegraphics[width=\marginparwidth]{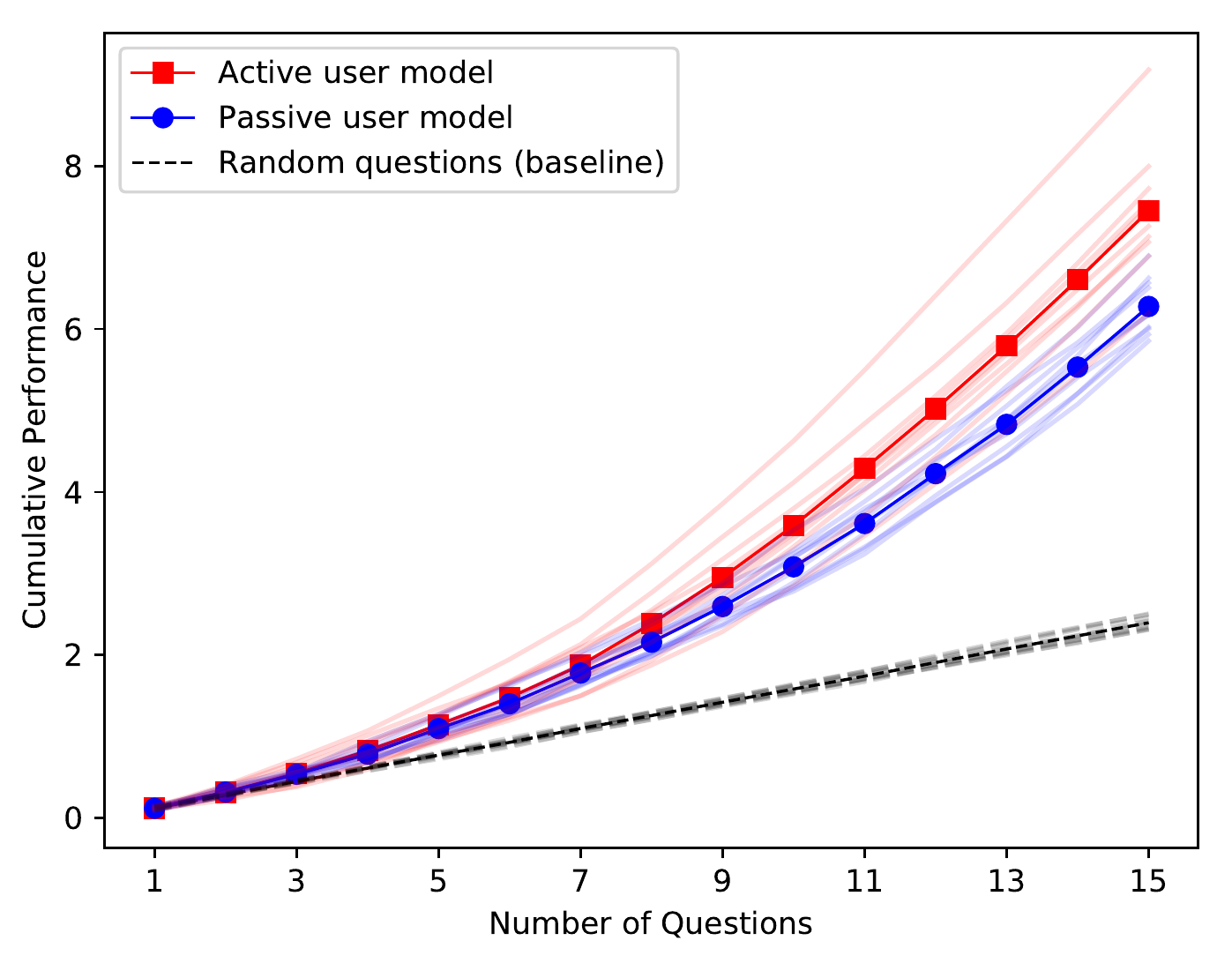}
\Description{User study results}
\caption{
Average cumulative reward curves for different conditions of the user study. Shaded lines show the average performance of individuals (over 20 target words) and solid lines show the average over the participants. The task performance consistently improved when interacting with the AI that had an active user model of the participants.
}
\label{fig:user_study_cum_reward_all}
\end{marginfigure}

We conducted a user study on 10 participants interacting with both AIs for 20 different target words. Figure~\ref{fig:user_study_cum_reward_all} depicts the performance of participants in the different conditions (individual performances are shown by shaded lines and averages over participants by solid lines). Participants achieved noticeably higher performance while interacting with the active user model (red) compared to the passive user model (blue). This difference was at a significant level after 12 questions (see \cite{peltola2018modelling} for details). The performance of a random question system is shown as a baseline.  

\section{Implications for human--computer interaction}

If better collaboration comes from understanding your collaborators better, and if understanding them comes from ToM capabilities, how can we endow an intelligent system with these capabilities? We believe this calls for a major paradigm shift. Current modes of human--AI interaction treat the users as sources of data, and assume users treat the system as a tool. If we are to understand users better, we need to formulate human--AI interaction as a multi-agent problem. This opens up the possibilities for richer interactions and also ways to incorporate ToM-based methods for modelling the user.

Conversely, the role of understandability and predictability of the system for the user becomes even more emphasised, as now not only the user experience, but also the statistical models underlying the system will depend on it. This calls for collaboration between researchers in human--computer interaction, cognitive science, explainable AI, and machine learning, in order to develop human-understandable intelligent learning systems for interactive use and to accommodate the relevant human factors in user models tailored for varying applications.

\bibliographystyle{ACM-Reference-Format}
\bibliography{references}

\end{document}